\documentclass[epj]{svjour}
\usepackage{graphics}
\usepackage{amssymb}
\usepackage{graphicx}
\usepackage{subfig}
\usepackage{amsmath}

\begin{document}
	\title{Developing a single phase liquid argon detector with SiPM readout}

	\author{L.~Wang\inst{1,2}, Y.~Lei\inst{3}, T.A.~Wang\inst{4}, C.~Guo\inst{1,2,5}		\thanks{\emph{Corresponding address:}  guocong@ihep.ac.cn}, K.K.~Zhao\inst{6}, X.H.~Liang\inst{2,7}\thanks{\emph{Corresponding address:}   liangxh@ihep.ac.cn}, S.B.~Wang\inst{1} \and Y.D.~Chen\inst{3} 
	}                     
	\institute{Experimental Physics Division, Institute of High Energy Physics, Chinese Academy of Sciences, Beijing, China \and School of Physics, University of Chinese Academy of Sciences, Beijing, China \and School of Electronic, Electrical Engineering and Physics, Fujian University of Technology, Fuzhou, China \and State Key Laboratory of High Power Semiconductor Laser, College of Physics, Changchun University of Science and Technology, Changchun, Jilin, China \and State Key Laboratory of Particle Detection and Electronics, Beijing, China \and School of Physics, Sichuan University, Chengdu, China \and Astro-particle Physics Division, Institute of High Energy Physics, Chinese Academy of Science, Beijing, China}
	\date{the date of receipt and acceptance should be inserted later}
	%
	\abstract{
	Liquid argon is used as a target material in several current and planned experiments related to dark matter direct searching and neutrino detection. SiPM is becoming the standard for scintillator detectors because of its advantages over traditional PMT. In this paper, we developed a single-phase liquid argon detector using eight 1 $\times$1 inch$^2$ Hamamatsu S14161-6050HS 4×4 SiPM arrays. The directly measured light yield is 25.7 $\pm$ 1.6 photo-electrons per keV, which corresponds to 12.8 $\pm$ 0.8 photo-electrons primarily generated by the argon scintillation. The rest is contributed by the cross-talk and after-pulse of SiPM. In addition, we provide an experimental method to estimate the effect of crosstalk and afterpulse on light yield using dark noise data. Finally, we quantitatively give the relationship between the light yield and the decay time of the slow component of a liquid argon detector.
		\PACS{
			{PACS-key}{discribing text of that key} 
		} 
	} 

	\authorrunning{L. Wang et al}
	\titlerunning{}
	\maketitle

	\section{Introduction}
	\label{intro}
	Liquid argon has been recognized as an extremely attractive target material in the field of dark matter direct detection~\cite{warp,DarkSide-20k,Deap2008} and neutrino detection~\cite{Gerda,DUNE1} because of its high light yield and excellent n/$\gamma$ discrimination~\cite{Deap2008,PSDpaper1,PSDpaper2,PSDpaper3}. The depleted argon, discovered by the Darkside collaboration~\cite{DarkSide-20k} made it more competitive in low background experiments. Silicon photomultiplier (SiPM), the next generation of photosensors, is becoming the standard for scintillation detectors~\cite{SiPM}. SiPM-based photon detection modules are also essential for liquid argon detectors, and the DS-20k collaboration~\cite{DS-SiPM} and the DUNE collaboration~\cite{DUNE-SiPM} are also developing SiPM-based photon detection modules for applications in liquid argon.
	
	The 1 $\times$1 inch$^2$ Hamamatsu S14161-6050HS 4×4 SiPM array is a good photosensor candidate for liquid argon detector and our previous works~\cite{VUV4_SiPM,SiPM_array} has demonstrated its stability operating at liquid argon temperatures. Using this kind of SiPM array, we developed a SiPM-based single-phase liquid argon detector and calibrated it with a $^{241}$Am $\gamma$ source. This paper is organized as follows. Sec.2 presents the detail of the experiment setup. Sec.3 describes the calibration results of the detector with an $^{241}$Am $\gamma$ source. Sec.4 introduces an experimental method to eliminate the overestimation of light yield because of SiPM crosstalk (CT) and afterpulse (AP) and then gives the corrected light yield result of the experiment. Sec. 5 discusses the relationship between the light yield and the purity of a liquid argon detector. And Sec.6 is the summary.
	
\section{Experimental setup}\label{sec:section2}

\subsection{Detector subsystem}

\begin{figure}[htb]
	\centering
	\includegraphics[width=7cm]{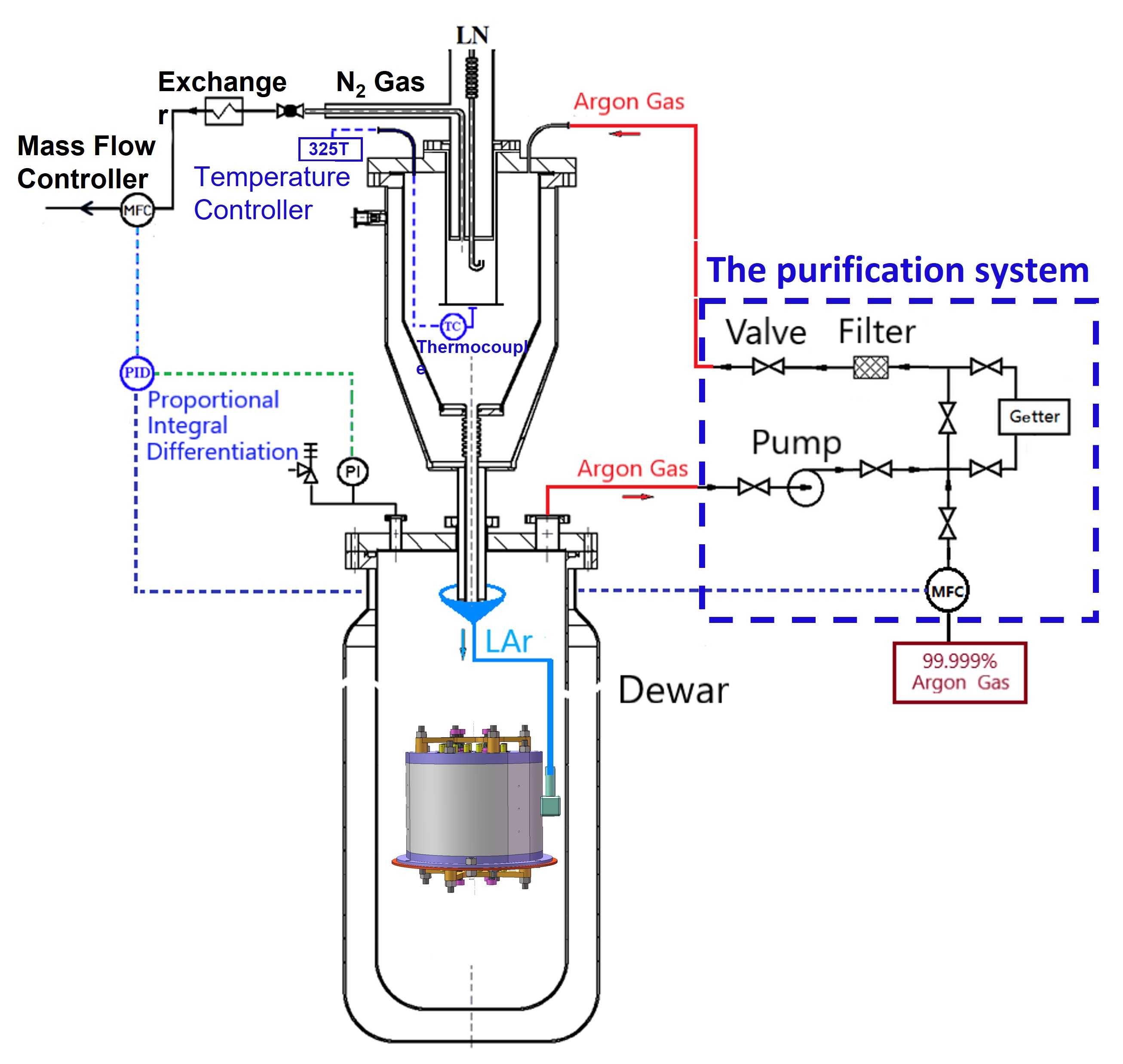}
	\caption{\label{detector} Schematic diagram of the LAr detector.}
\end{figure}
\begin{figure}[!h]
	\centering
	\begin{minipage}[b]{0.45\linewidth}
		\subfloat[Four SiPM arrays on the top.]{\label{SiPM1}
			\includegraphics[width=1.4in]{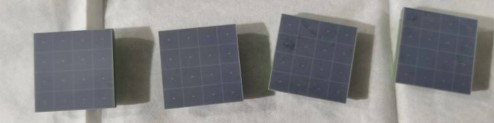}} \\  
		\subfloat[Four SiPM arrays on the bottom.]{\label{SiPM2}
			\includegraphics[width=1.4in]{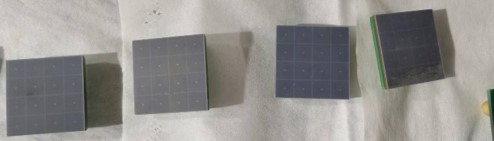}} \\  
		\subfloat[A bottom view of the detector. SiPM arrays at the bottom had not been assembled yet.]{\label{detector1}
			\includegraphics[width=1.4in]{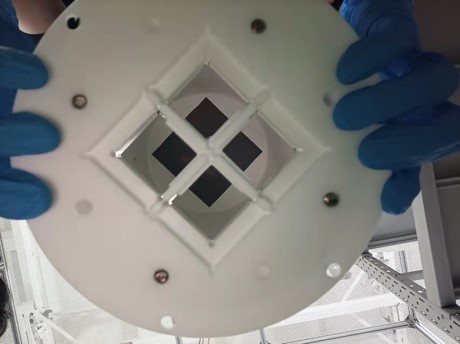}}
	\end{minipage} 
	\begin{minipage}[b]{0.45\linewidth}
		\centering
		\subfloat[The center detector. It would be immersed in liquid argon directly.]{\label{detector2}
			\includegraphics[width=1.4in]{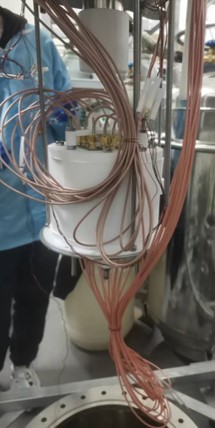}}
	\end{minipage}
	\caption{\label{Photographs}}
\end{figure}

 
Fig.~\ref{detector} shows the schematic diagram of the liquid argon detector. A liquid nitrogen based refrigeration system is used to liquefy 99.999\% high-purity argon gas and the liquified argon is contained in a dewar. A polytetrafluoroethylene (PTFE) sleeve with a diameter of 76~mm and a height of 76~mm is placed inside the dewar as the main component of the center detector.  Eight 1$\times$1 inch$^2$ Hamamatsu S14161-6050HS 4×4 SiPM arrays are placed on the top and bottom of the sleeve to serve as photo-sensors. Several photographs have been shown in Fig.~\ref{Photographs}. The entire inner surface of the cylinder, except for the photocathode of SiPM, is covered by enhanced specular reflector film (ESR) to enhance the light collection. To improve the photon detection efficiency, both the SiPM surface and the ESR surface are coated with 1,1,4,4-tetraphenyl1-1,3-butadiene (TPB) to convert the 128 nm photons to 420 nm. The thickness of the TPB is $\sim$200~$\mu$g/cm$^2$. An $^{241}$Am $\gamma$ source is put in the center of the sleeve with a copper wire. The $^{241}$Am was electro-plated on one side of a stainless steel disc with a diameter of $\sim$6~mm. Given that the electronegative impurities in liquid argon reduce the light yield of the detector~\cite{NO-LAr,N-LAr}, a getter, which was made by Beijing Beiyang United Gas Co., Ltd, is used in the experiment to remove impurities such as nitrogen, oxygen, and water vapor. The newly liquefied argon is delivered to the inside of the detector's PTFE cylinder to ensure that the liquid argon inside the detector has the highest purity. The details of the gas circulation and purification system can be found in Ref~\cite{Peixian}.

\subsection{FABs and electronics}

\begin{figure}[htb]
	\centering
	\includegraphics[width=8cm]{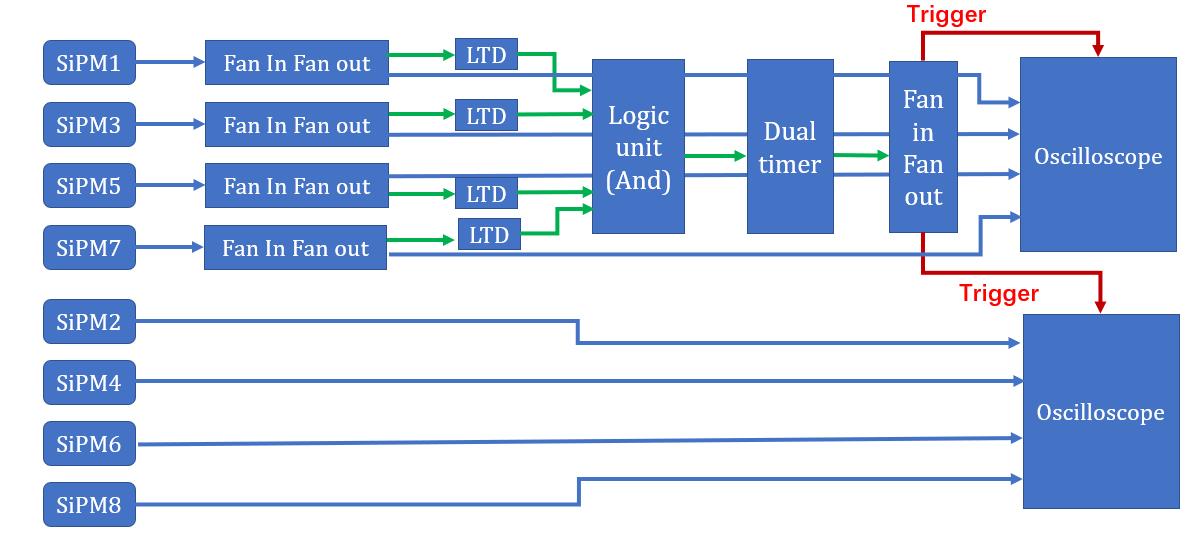}
	\caption{\label{readout} Readout diagram of the LAr detector.}
\end{figure} 

Front-end amplifier circuit boards (FABs) were designed for pre-amplifying signals from SiPM arrays. Our previous work has proved its stability at a low temperature down to 77~Kevin. Two DH 1765-4 DC power suppliers were used for bias supplying for eight SiPM arrays and two RAGOL DP831A DC power supplies were used to power up eight FABs. More details about the FABs could be found in Ref.~\cite{VUV4_SiPM,SiPM_array}.

Fig.~\ref{readout} shows the data flow of the electronics. SiPM 1, 2, 3, and 4 were placed on top of the detector. On the contrary, SiPM 5, 6, 7, and 8 were placed at the bottom. For collecting signals from eight SiPM arrays spontaneously, two LeCroy 610Zi oscilloscopes were used. The trigger was set as the coincidence of SiPM 1, 3, 5, and 7. The thresholds of them were set to about 2.5~p.e. 

\section{Experimental results}\label{sec:section3}
\subsection{Single photoelectron calibration}

\begin{figure}[htb]
	\centering
	\includegraphics[width=8cm]{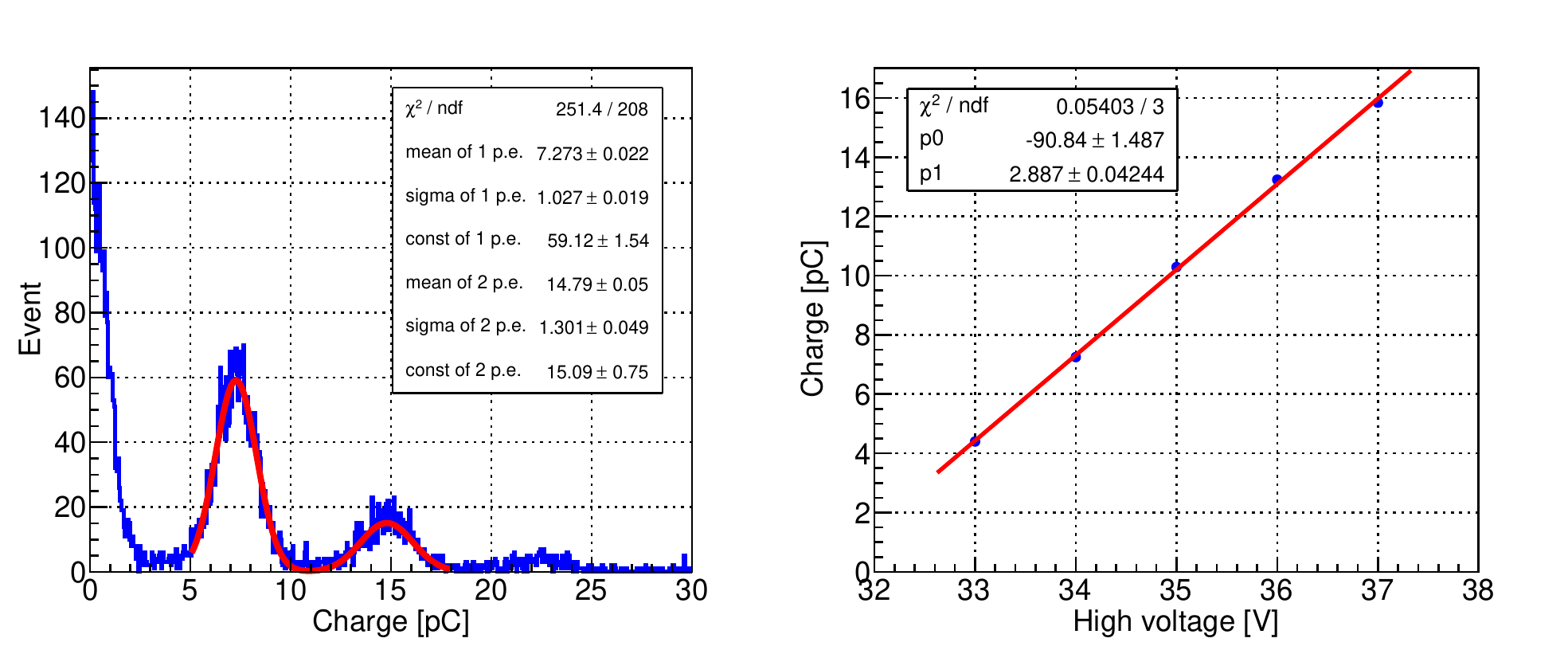}
	\caption{\label{SPE} Left: The p.e. distribution of SiPM 1 at V$_{bias}$ = 34.0~V. Right: The calibration result of SiPM 1 with different V$_{bias}$ from 33.0~V to 37.0~V. }
\end{figure} 

\begin{figure}[htb]
	\centering
	\includegraphics[width=8cm]{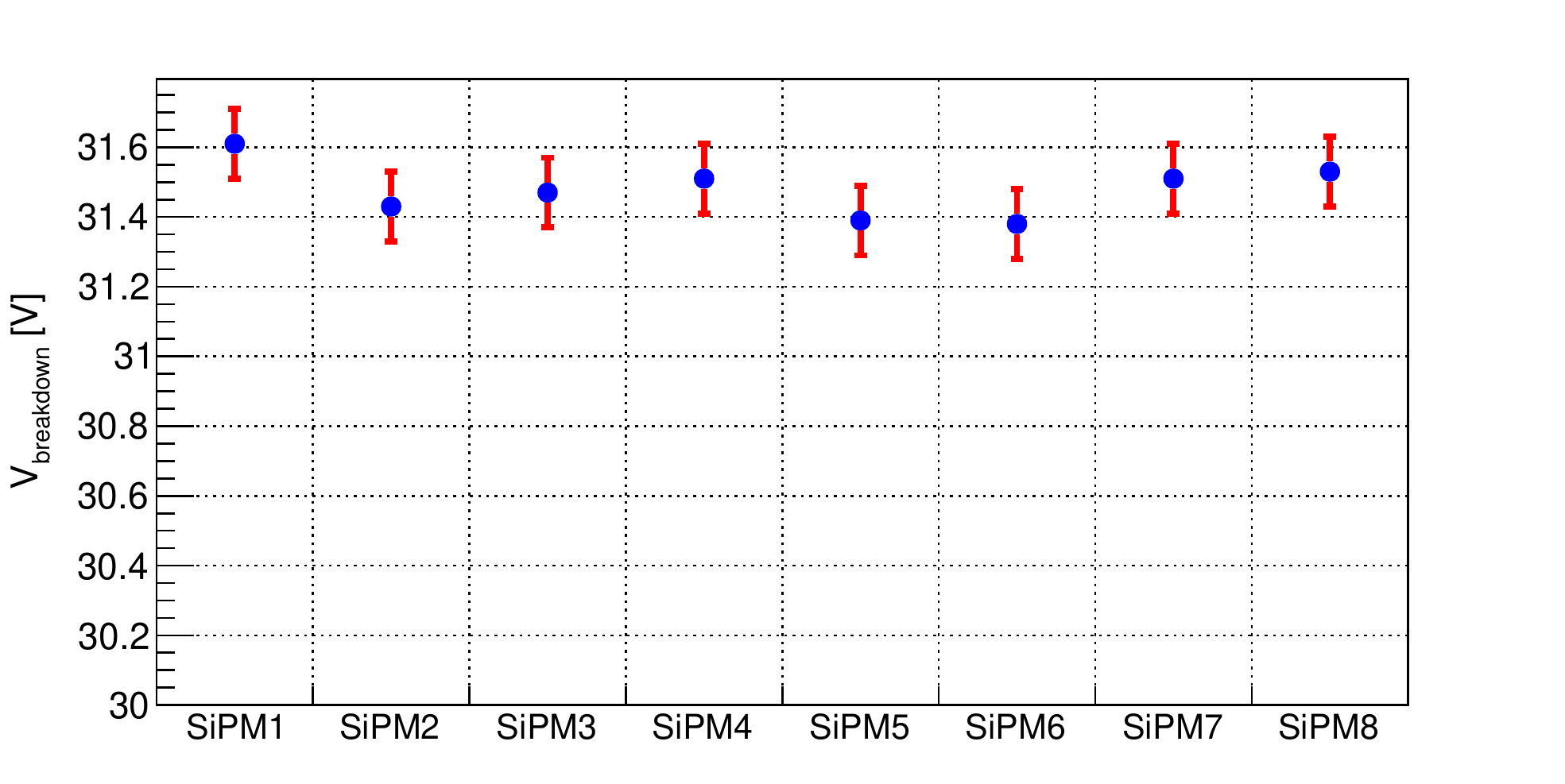}
	\caption{\label{LY_Graph} V$_{breakdown}$ distribution of all eight SiPM arrays. The highest point is 0.7$\%$ higher than the lowest point.}
\end{figure} 

Eight SiPM arrays were calibrated with a LED, and the LED is placed on the upper side of the interior of the dewar, which is at room temperature. The histogram in the left side of Fig.~\ref{SPE} shows the LED spectrum of SiPM 1 at V$_{bias}$ = 34.0~V. The charge of 1~p.e. could be estimated by the difference of 2 p.e.~peak and 1 p.e.~peak. And the plot on the right side of Fig.~\ref{SPE} gives the calibration result of SiPM 1 with different V$_{bias}$ from 33.0~V to 37.0~V. A linear function was used to fit the points to evaluate the zero-gain voltage V$_{breakdown}$, which is a specific voltage where the gain of the SiPM array drops down to zero and is defined as the X-intercept of the fitting function. Usually, V$_{breakdown}$ is slightly different for various SiPM arrays, although they are made by the same vendor. According to the fitting results, V$_{breakdown}$ of all eight SiPM arrays were presented in Fig.~\ref{LY_Graph}.

\subsection{$^{241}$Am calibration}

\begin{figure}[htb]
	\centering
	\includegraphics[width=9cm]{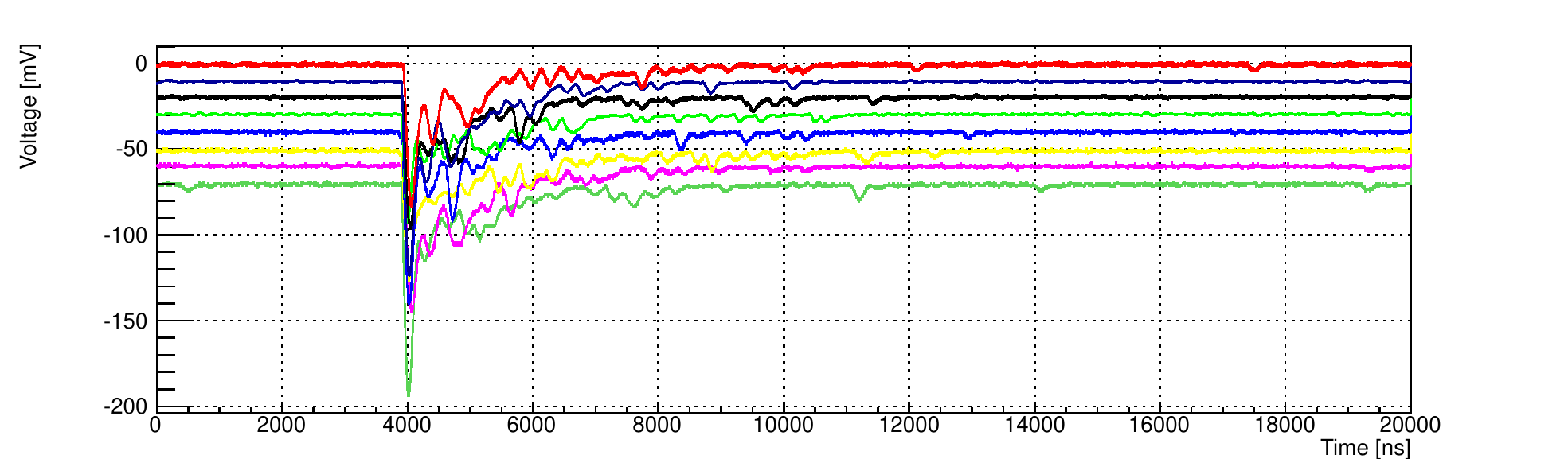}
	\caption{\label{SamplePulse} An example of eight SiPM output pulses created by a $\gamma$-ray.}
\end{figure} 

\begin{figure}[htb]
	\centering
	\includegraphics[width=9cm]{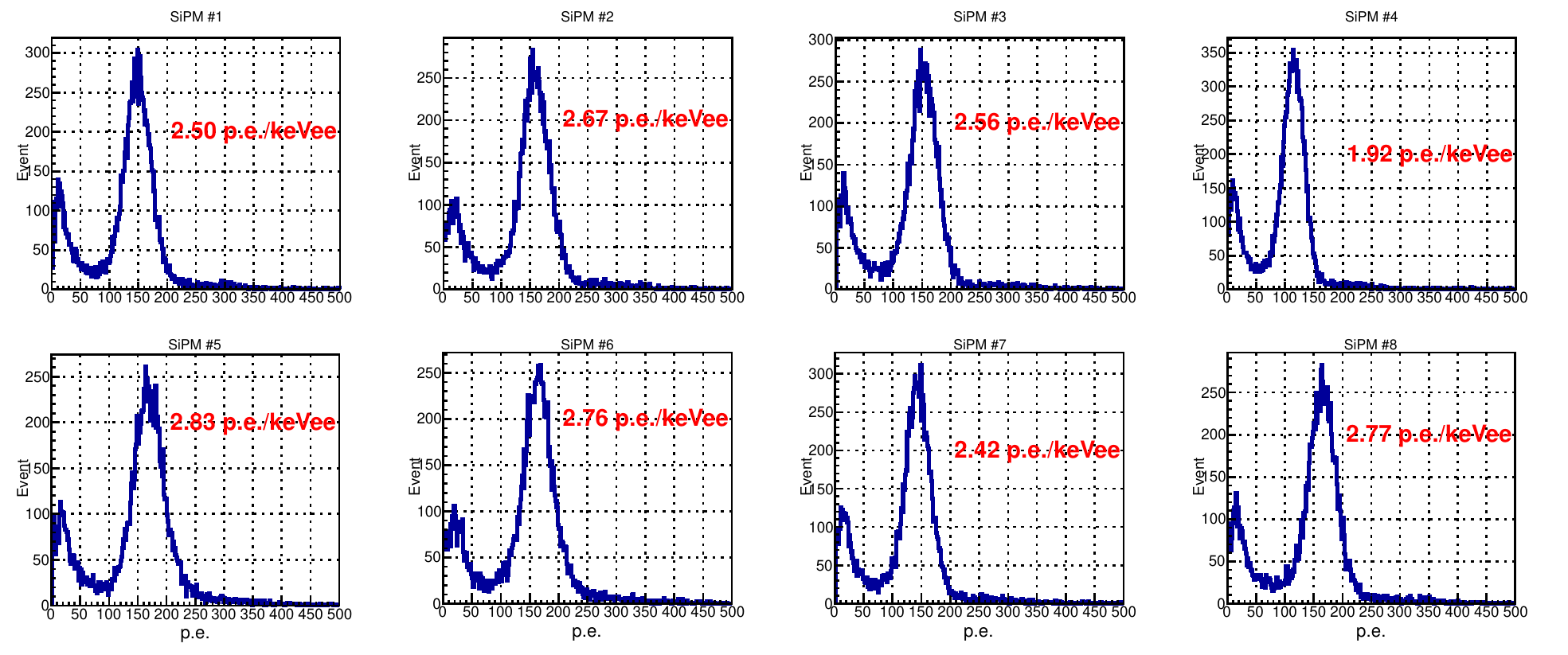}
	\caption{\label{8SiPM}Energy responses of all eight SiPM arrays with a $^{241}$Am source.}
\end{figure}

Fig.~\ref{SamplePulse} shows an example of $^{241}$Am $\gamma$-ray pulses presented from eight SiPM arrays. The trigger was set at 4000~ns and the time window was 20000~ns. The individual $^{241}$Am energy spectra of the eight SiPM arrays are shown in Fig.~\ref{8SiPM}. The bias voltages of the SiPMs were set to 36~V, which corresponds to V$_{over}\approx$ 4.5~V. The biggest peaks in the eight spectra were the full-energy peaks from the $^{241}$Am 59.6~keV $\gamma$-ray.  Their light yields were marked out in red in each histogram separately.  The light yield differences are mainly caused by the V$_{breakdown}$ differences and TPB thickness differences of the SiPM arrays, while the much smaller light yield of SiPM 4 is due to the light blocking of the $^{241}$Am source.

\begin{figure}[!h]
	\centering
	\includegraphics[width=9cm]{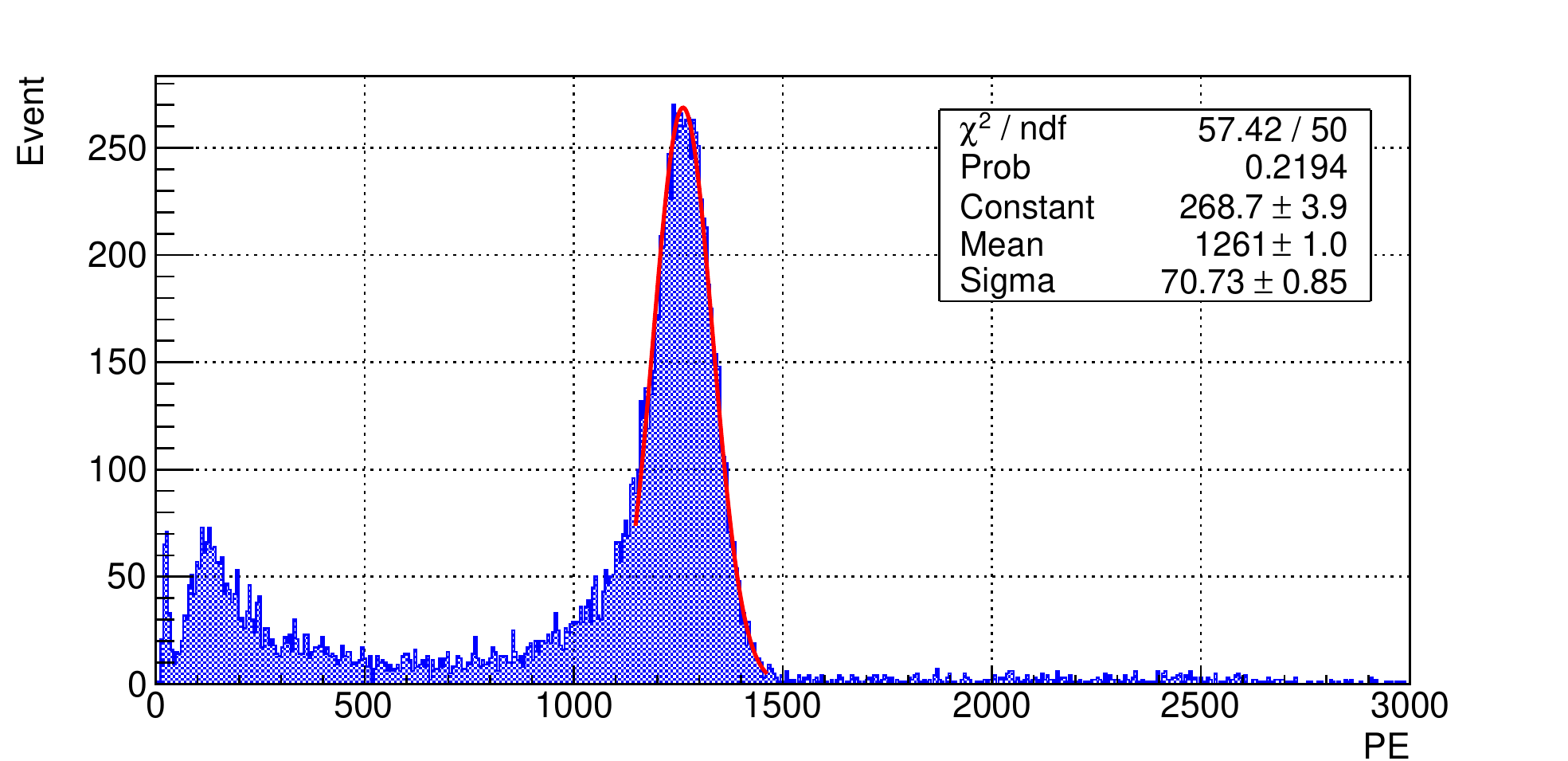}
	\caption{\label{Am241Spectrum} $^{241}$Am energy response of the LAr detector. the Gaussian fitted result of the full-energy peak gives the light yield of 21.14$\pm$0.02 p.e./keVee. }
\end{figure} 

Combining data of eight channels, the $^{241}$Am response of the liquid argon detector is plotted in Fig.~\ref{Am241Spectrum}. The 59.6~keV $\gamma$-ray peak was fitted by a red curve according to a Gaussian function. The light yield at V$_{bias}$ = 36.0~V is 21.14$\pm$0.02 p.e./keVee, with the energy resolution of 5.61$\pm$0.01$\%$.

\section{Light yield correction} \label{sec:section4}
\subsection{Estimation of contributions from crosstalks and afterpulses}

\begin{figure*}[htb]
	\centering
	\includegraphics[width=13.5cm]{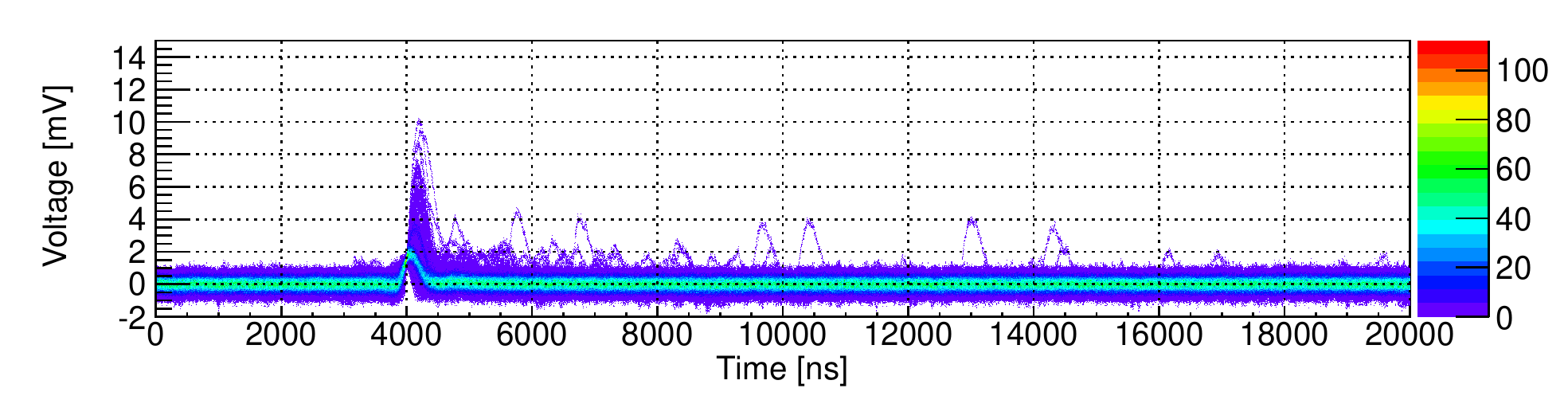}
	\qquad
	\includegraphics[width=13.5cm]{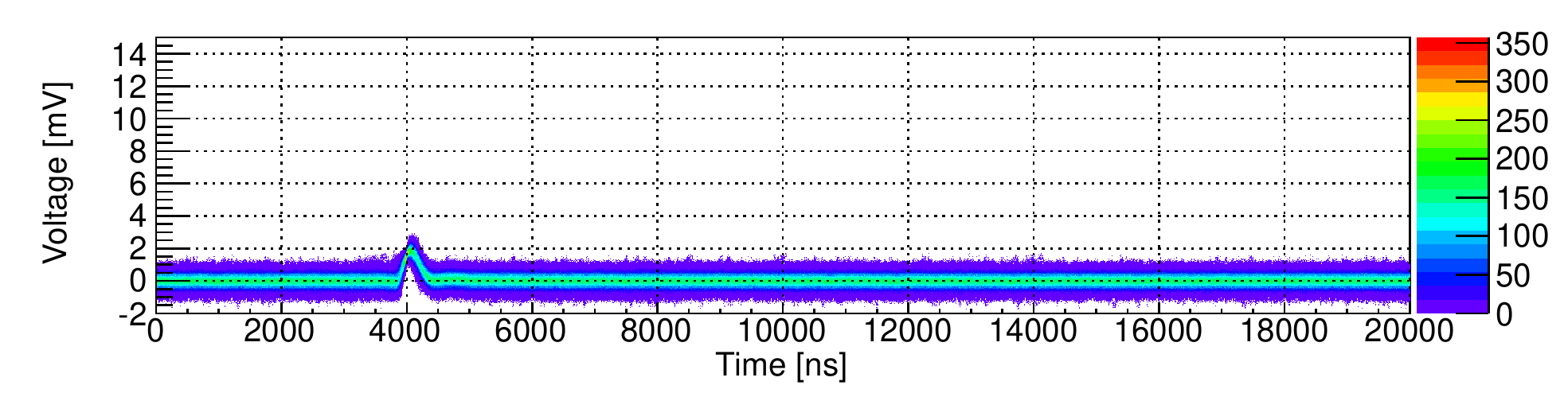}
	\caption{\label{CToverlap} Top: an overlapping profile of thousands of dark signals collected by self-trigger in a dark environment. lots of cross-talk like direct cross-talk, delay cross-talk and after pulses could observe as a part of primary single photo electron pulses. Bottom: a set of dark signals artificially cutting out events that do not include CT parts.} 	
\end{figure*}

\begin{figure}[htb]
	\centering
	\includegraphics[width=9cm]{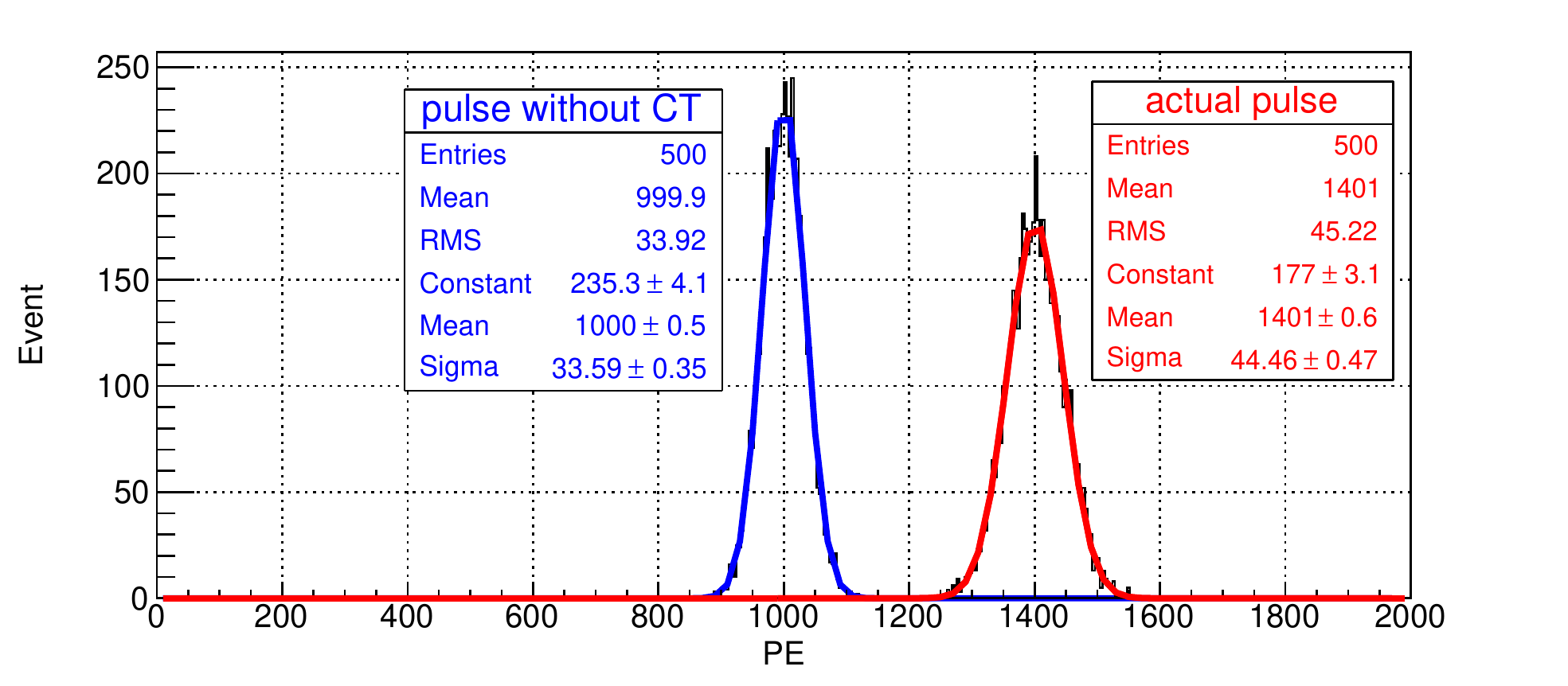}
	\caption{\label{CT} Simulate total p.e.~number distribution with~(red) and without~(blue) cross talks for 35.0~V bias voltage. A conversion factor of 1.4 is found to correct the measured light yield to the true one. } 	
\end{figure}

The optical crosstalks of SiPM are the result of the second electron avalanche in adjacent pixels. And afterpulses are derived from the avalanche of trapped electron in the same pixel. In experiment, crosstalks and afterpulses were believed to result in a false exaggeration in the light yield estimation, since they are not actually the light emission from the crystal. 

The contributions from crosstalks and afterpulses of SiPM arrays should be eliminated during the data analysis. And our previous work indicated that unlike PMT, the probabilities of crosstalk and afterpulsing in SiPM are not negligible~\cite{SiPM_array}. Thus, a Monte Carlo simulation method was developed below to estimate the contributions of crosstalks and afterpulses.

In this section, we introduce the method based on dark noise data to quantitatively estimate the contributions from SiPM crosstalks and afterpulses. The principle of this method is to simulate the total p.e.~spectrum of $\gamma$ peaks using waveforms with and without crosstalks and afterpulses respectively. The waveform sample with crosstalks and afterpulses, shown in the top panel of Fig.~\ref{CToverlap}, is a collection of thousands of dark noise signals of the SiPMs. Substantial crosstalks and afterpulses could be observed compared with the selected clean waveforms in the bottom panel. The clean waveforms are selected by throwing the waveforms with crosstalks and afterpulses away. The cuts used here are the highest amplitude of each primary pulse and the smoothness of the baseline after each primary pulse.

In the toy MC, the true p.e.~number is set to 1000. Selecting randomly 1000 waveforms from the top and bottom panels and converting the total charge to reconstructed p.e.~number, the ratio of the mean values of the two p.e.~spectra is the overestimate of the true p.e.~number due to crosstalks and afterpulses. Figure~$\ref{CT}$ gives the result of the simulation at a bias voltage of 35.0~V. Comparing the means of the two histograms, a factor of 1.4 is the p.e. overestimate for this bias voltage. Since the crosstalks and afterpulses probabilities change along with the bias voltages, the correction factors were estimated at different V$_{bias}$, and the numbers are 1.4, 1.7, and 2.0 for bias voltages 35.0~V, 36.0~V, and 37.0~V, respectively.

\subsection{Light yield from liquid argon scintillation}

\begin{figure}[!htp]
	\centering
	\includegraphics[width=9cm]{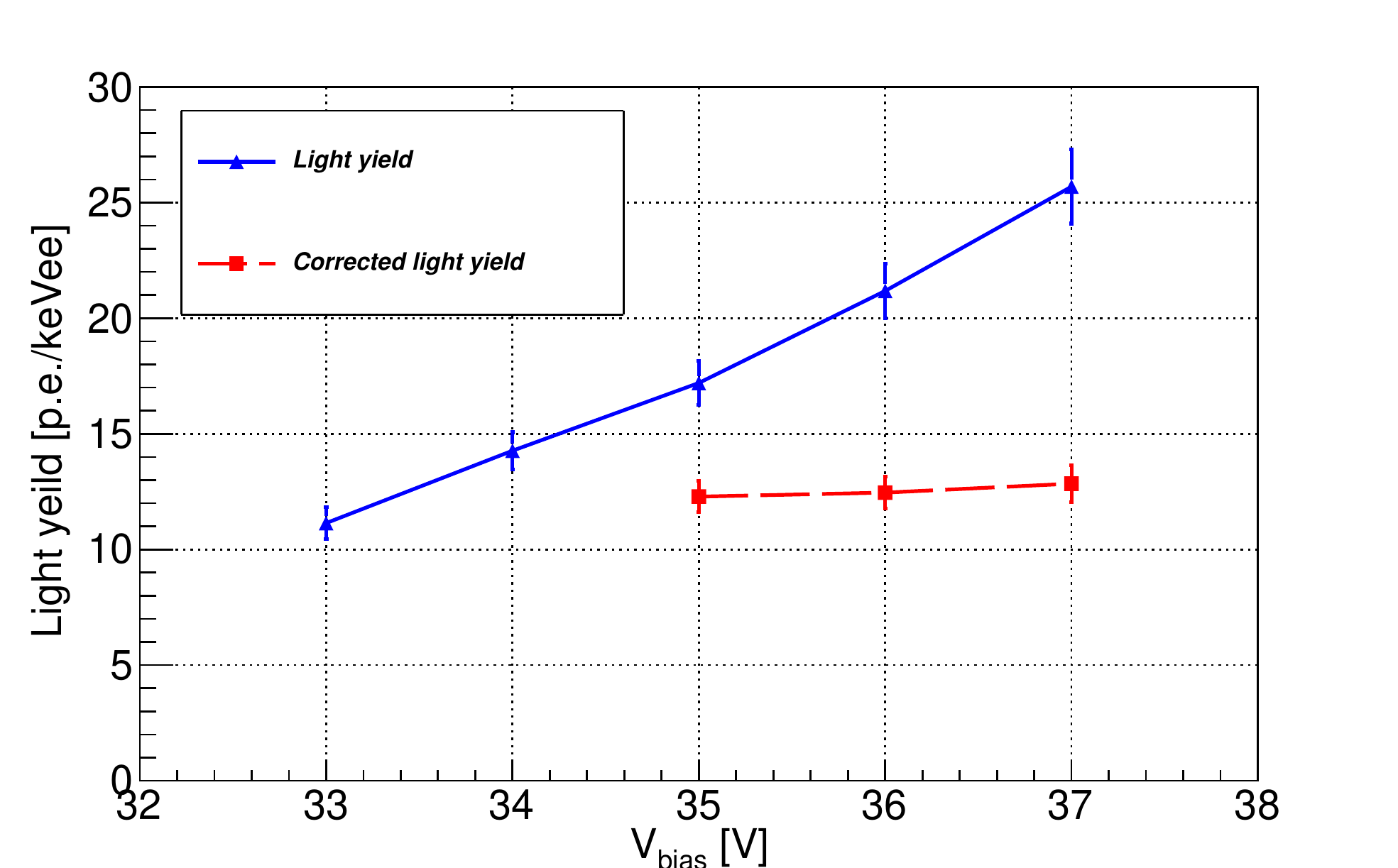}
	\caption{\label{lightY} The blue solid line corresponds to the light yield of the liquid argon detector with eight SiPM arrays. It is a monotone increasing function of V$_{bias}$ as a result of the increasing CT probability. After cutting off CT, the tendency of corrected light yield with V$_{bias}$ is calculated as the red dotted line.} 	
\end{figure}

Unlike the photo-multiplier tubes, the light yield measured by the SiPM array is not constant with various V$_{bias}$. Typically, it is noticed that the increase of V$_{bias}$ would enhance the photon detection efficiency of a SiPM array~\cite{HamaDS}. Besides, the crosstalks and afterpulses probabilities are also positively related with V$_{bias}$. 

Fig.~\ref{lightY} shows the light yield function with V$_{bias}$ as a solid blue line. At V$_{bias}$=37.0~V, the light yield reaches up to 25.7$\pm$1.6 p.e./keVee, according to the data of eight SiPM arrays. However, based on the estimation discussed in sec.~\ref{sec:section4}, a correction factor of 2 should be introduced at 37.0~V. The corrected result by eliminating the effects of crosstalks and afterpulses was shown as a red dotted line in Fig.~\ref{lightY}. After correction, a light yield of 12.8 $\pm$ 0.8 p.e./keVee at 37.0~V is obtained, which is generated by the direct liquid argon scintillation. 

\section{Relationship between light yield and liquid argon purity}\label{sec:section6}
The scintillation process of liquid argon has been experimentally established and theoretically well-understood~\cite{PRB1978,PP1981}. The impurities in liquid argon cause a decrease in the light yield of liquid argon detectors is a well-known phenomenon, while accurate measurements are rare. Liquid argon luminescence theoretically has a fast component with a decay time of 7~ns and a slow component with a decay time of 1600~ns and the impurities in liquid argon affect the decay times~\cite{NPB2009}. Considering the photosensor's timely response, the slow component's decay time is often used in experiments to characterize the impurity content inside the detector. This section discusses the relationship between light yield and decay time of the slow component $\tau_{slow}$.

\subsection{Pulse fitting}

\begin{figure}[htb]
	\centering
	\includegraphics[width=9cm]{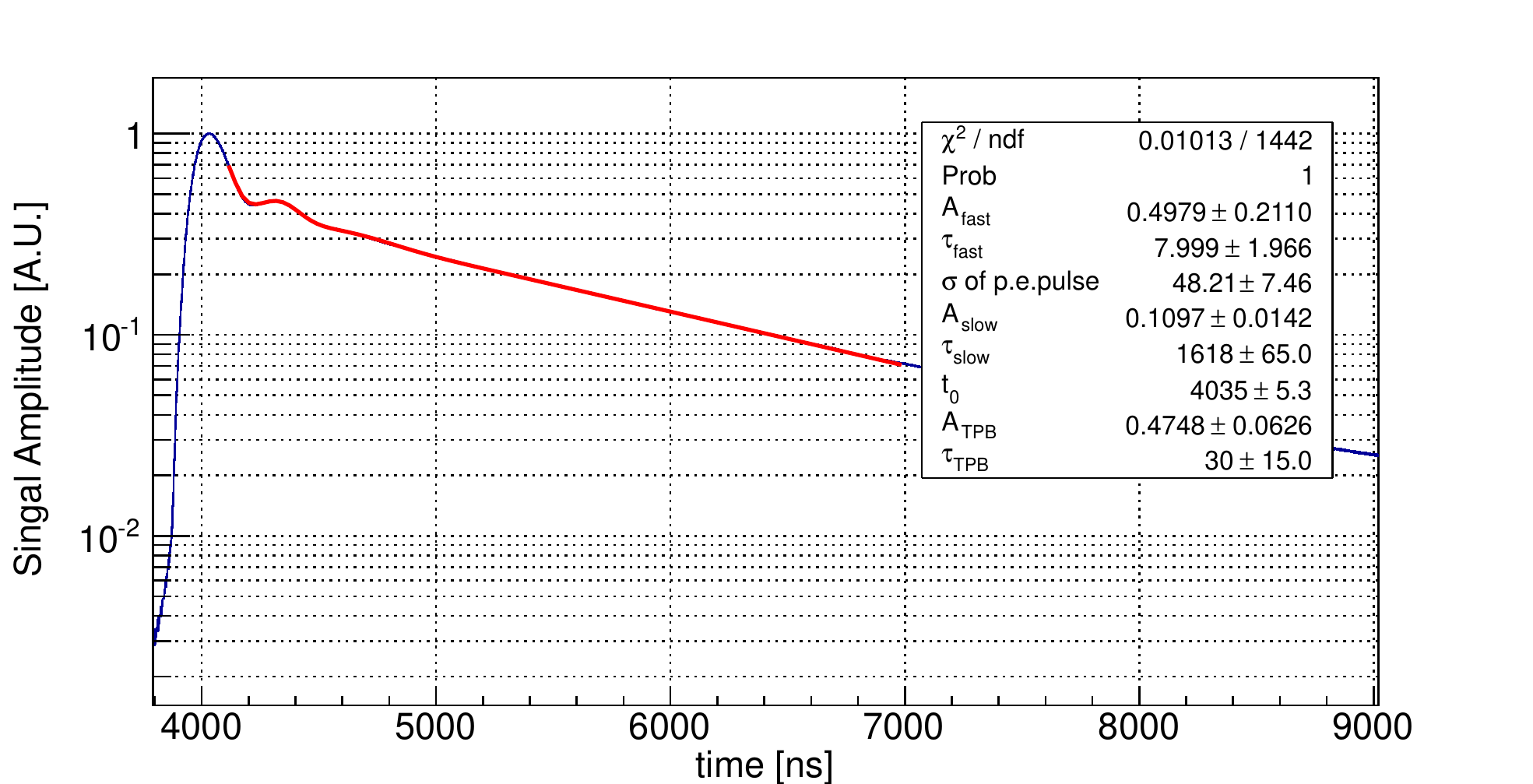}
	\caption{\label{pulse} The normalized average waveform for $\gamma$-ray. It is fitted by Eq.~\ref{fittingfunction}. The decay time of the slow component is 1618$\pm$65~ns.}
\end{figure}

The decay times of the slow components were obtained by fitting the normalized average waveforms according to Eq.~\ref{fittingfunction}. The fitting function mainly consists of two parts. The first part is $I(t)$ (Eq.~\ref{eq1})~\cite{TPB2}, which includes three exponentials. They are the fast component and slow component based on the principle of liquid argon luminescence, and another exponential corresponding to the TPB scintillation process. Several research indicated the existence of the TPB component and claimed it as a significant part of the liquid argon pulse shape analysis~\cite{TPB1,TPB2}. Then, the basic pulse shape function could be given by the sum of convolutions of each exponential function with one Gaussian function. The Gaussian component represents the pulse shape of SiPM single p.e. 

The second part of Eq.~\ref{fittingfunction} is two Gaussian functions with equal interval $t_1$. The presence of this part is the result of our FAB design defect, which we believe is likely to be the unmatching of capacitors between the FAB and SiPM array in cryogenic environment. Because of that, several reflected bumps follow behind the primary pulse at a $\sim$300~ns interval. Bumps could be observed easily in Fig.\ref{pulse}. The bump amplitude is about 10 times weak in each reflection. Here we only fitted the first two bumps because bumps after more than two reflections are too weak to affect the average pulse shape.

	\begin{equation}
		\begin{aligned}
		I(t) = \sum_{j =  fast,slow,TPB } \frac{2 A_{j}}{\tau_{j}} \exp \left[\frac{\sigma^{2}}{2 \tau_{j}^{2}}-\frac{t-t_{0}}{\tau_{j}}\right]
		\\
		 \cdot\left(1-{Erf}\left[\frac{\sigma^{2}-\tau_{j}\left(t-t_{0}\right)}{\sqrt{2} \sigma \tau_{j}}\right]\right) 
		\label{eq1}
		\end{aligned}
	\end{equation}

	\begin{equation}
		\begin{aligned}	
		F(t) = I(t)+A\cdot Gaussian(t-t_{1})
		\\
		+B\cdot Gaussian(t-2\cdot t_{1})
		\label{fittingfunction}
		\end{aligned}
	\end{equation}

Fig.\ref{pulse} shows the fitting result based on Eq.~\ref{fittingfunction}. Important parameters were listed in the legend. $A_{fast}$, $A_{slow}$ and $A_{TPB}$ are the amplitudes of three exponentials separately. $\tau_{fast}$, $\tau_{slow}$ and $\tau_{TPB}$ are the decay times. $\sigma$ of p.e. pulse means the $\sigma$ of a typical single photoelectron pulse generated by the SiPM array. t$_0$ is the start time of the pulse. Besides, t$_{1}$ in Eq.~\ref{fittingfunction} is the time difference between the first reflection and the primary pulse.

\subsection{Results}

\begin{figure*}[htb]
	\centering
	\includegraphics[width=12cm]{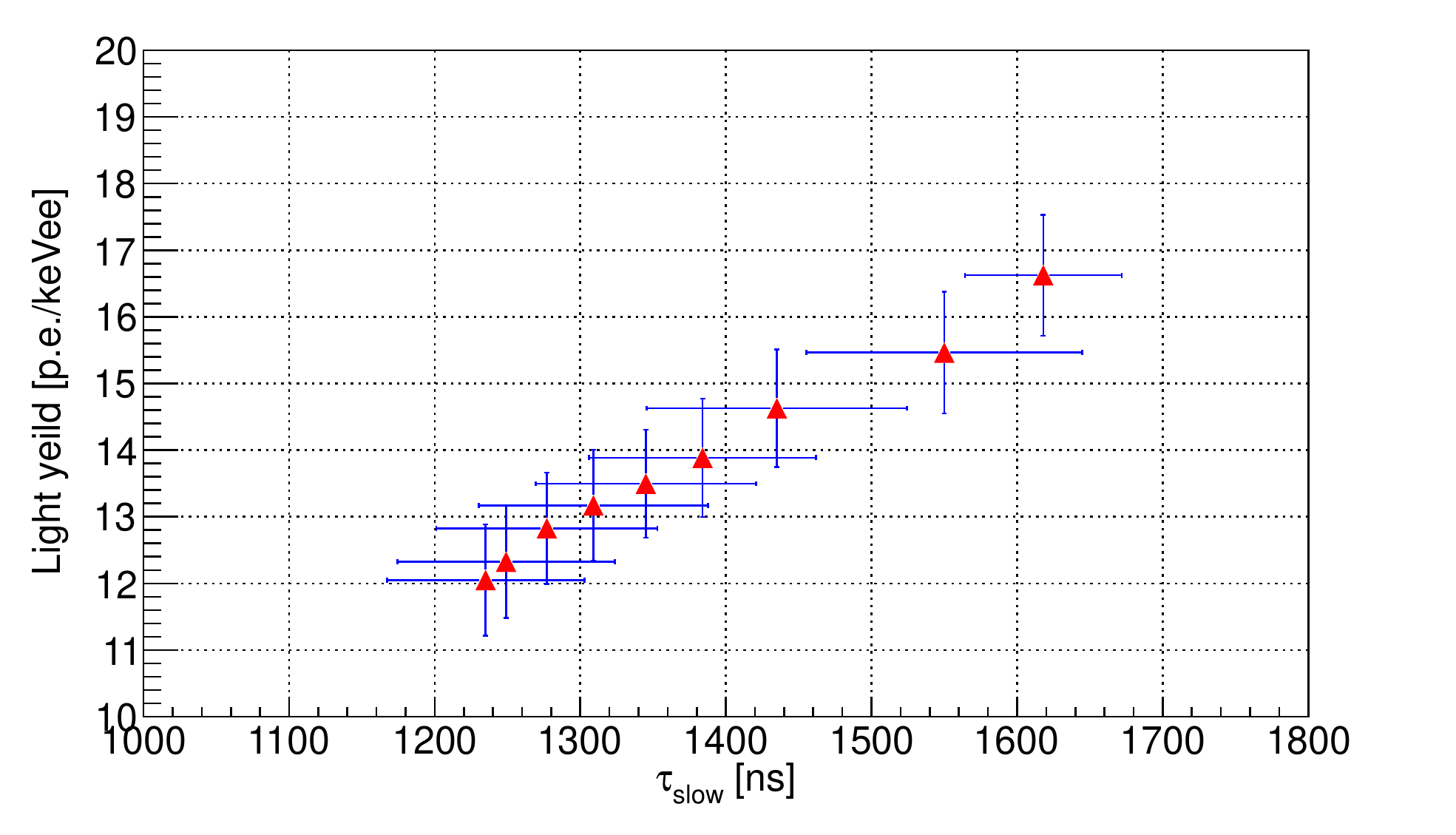}
	\caption{\label{Tau} Light yield of the liquid argon detector with different slow components $\tau_{slow}$. The error of the x-axis is the fitting uncertainty of Eq.~\ref{fittingfunction}. And the error of the y-axis is the statistic uncertainty of light yield.}
\end{figure*}

While measuring the relationship between the light yield and the purity of a liquid argon detector, the SiPM arrays are operated at V$_{bias}$ = 35~V. And the decay time $\tau_{slow}$ was used to characterize the purity of liquid argon. As a result, the function of the liquid argon light yield with $\tau_{slow}$ was estimated accurately. The result is given in Fig.~\ref{Tau}. The enhancement of the light yield during argon purifying was demonstrated quantitatively. For every 100ns increase in decay time, the light yield of the liquid argon detector increases by $\sim$10\% in the $\tau_{slow}$ range from 1240~ns to 1618~ns.

\section{Summary}

A single-phase liquid argon detector has been developed with eight 1 $\times$1 inch$^2$ Hamamatsu S14161-6050HS 4×4 SiPM arrays and the detector has been stably operated for more than 50 days. An $^{241}$Am $\gamma$ source is used to calibrate the detector. The measured light yield is contributed by liquid argon luminescence and SiPM correlated signals. An experimental method based on dark noise data has been proposed to eliminate the contribution from SiPM crosstalks and afterpulses. A net light yield of 12.8 $\pm$ 0.8 p.e./keVee is obtained, which is one of the best light yield results for a liquid argon detector. 

Using the decay time of the slow component to characterize the purity of a liquid argon detector, we studied the relationship between light yield and liquid argon purity. With the increase of the decay time of the slow component, the light yield of a liquid argon detector increases, for every 100~ns improvement in decay time, the light yield increases by $\sim$10\%.

\section{Acknowledgments}

This work is supported by the National Key R \& D Program of China (Grant No. 2016YFA0400304) and the National Natural Science Foundation of China (Grant No. 12275289, Grant No.11975257, and Grant No.12175247).

	%
	%

\end{document}